\documentstyle[prl,aps,twocolumn]{revtex}
\begin{document}
\draft
\title{Gauge Theory of Gravity Requires Massive Torsion Field}
\author{Rainer W. K\"uhne}
\address{Fachbereich Physik, Universit\"at Wuppertal, 42097 Wuppertal, 
 Germany, kuehne@theorie.physik.uni-wuppertal.de}

\maketitle

\begin{abstract}
One of the greatest unsolved issues of the physics of this century is 
to find a quantum field theory of gravity. According to a vast amount 
of literature unification of quantum field theory and gravitation requires 
a gauge theory of gravity which includes torsion and an associated spin 
field. Various models including either massive or massless torsion fields 
have been suggested. We present arguments for a massive torsion field, 
where the probable rest mass of the corresponding spin three gauge boson 
is the Planck mass.
\end{abstract}

The connection used by general relativity [1] is symmetric. After 
Eddington [2] suggested to generalize general relativity by introducing 
an asymmetric connection, Cartan [3] associated angular momentum with the 
antisymmetric part (= torsion) of an asymmetric connection. The 
introduction of quantum mechanics [4] required a quantum theory of gravity 
whose quantities are no longer classical, but operators. After Yang 
and Mills [5] suggested to describe quantum field theories by gauge 
theories, Kibble [6] and Sciama [7, 8] attempted to describe gravity 
by a gauge theory, where they associated intrinsic spin [9] with Cartan's 
torsion. The successful description of the quantum field theory of the 
electroweak interaction by a spontaneously broken gauge theory [10, 11] 
and the subsequent proof for gauge theories to be renormalizable [12 - 16], 
inspired an increasing number of theorists to further develop gauge 
theories of gravity (for reviews see [17, 18]). We will briefly review the 
arguments for the need for a gauge theory of gravity and the need for 
a torsion field which we will show to be massive.

Classical electrodynamics and general relativity have well-known 
analogies. Resting electric charges are the sources of the static Coulomb 
field and rotating electric charges generate an extra magnetic field and 
an associated Lorentz force. The field equations of classical 
electrodynamics are the Maxwell equations, where the matter-free 
equations describe electromagnetic waves. By analogy, resting masses are 
the sources of the static gravitational field and rotating masses 
generate an extra gravitational field associated with the recently 
discovered [19] Lense-Thirring effect [20]. The field equations of general 
relativity are the Einstein field equations, where the linearized 
matter-free equations describe gravitational waves.

But there are also well-known differences. Electrodynamics can be 
quantized and the Maxwell equations remain the field equations of 
quantum electrodynamics. Quantization and renormalization are possible, 
because (in rationalized units) the Lagrangian has dimension $-4$ and 
the coupling constant dimension zero. By contrast, general relativity 
cannot easily be quantized, because the Lagrangian has dimension $-2$ 
and the coupling constant (Newton's constant) has dimension 2. Hence, a 
quantum version of general relativity is not renormalizable.

The aim is to find a quantum theory of gravity. Quantum field theories 
have to yield finite results for all orders of perturbation theory. 
Infinite contributions have to cancel one another via 
renormalization. The only quantum field theories yet known to be 
renormalizable are gauge theories [12 - 16].

Hence, the aim is to find a (quantum) gauge field theory of gravity. 
The first step is to find the appropriate gauge group.

The group underlying special relativity is the Poincar\'e group. Since 
general relativity is locally Lorentz invariant, the Poincar\'e group 
is a candidate for the gauge group underlying the gauge theory of 
gravity [6 - 8, 17].

The translational part of the Poincar\'e group is associated with the 
energy-momentum tensor and therefore with mass. As the metric tensor 
is of rank two, the gauge boson (``graviton'') associated with mass has 
intrinsic spin two.

The rotational part of the Poincar\'e group is associated with  
angular momentum [6 - 8, 17]. As the torsion tensor 
[3] is of rank three, its associated gauge boson (``tordion'' [17]) has 
intrinsic spin three.

The Einstein field equations are symmetric and can describe only spinless 
matter. This is because intrinsic spin is antisymmetric. The description 
of a Dirac field (which has spin $\hbar /2$) requires the introduction 
of torsion (which is antisymmetric) [6 - 8, 17].

The need for torsion and its association with angular momentum can be 
seen as follows. The Maxwell equations do not describe electricity and 
magnetism equivalently. An equivalent description requires the 
introduction of magnetic charges [21], where the $U(1)$ group of quantum 
electrodynamics is extended to the $U(1)\times U'(1)$ group [22]. The 
associated gauge bosons are the photon and Salam's ``magnetic photon'' 
[23]. By analogy, general relativity does not describe the translational 
part and the rotational part of the Poincar\'e group equivalently. 
An equivalent description requires the introduction of torsion (in analogy 
to magnetic charge). Furthermore, from the analogy between the Thirring-Lense 
effect and the Lorentz force we can infer the analogy between angular 
momentum and magnetic charge. Hence, both torsion and angular momentum 
are analogous to magnetic charge and therefore associated with one another.

The effects of orbital angular momentum are already described by 
general relativity (Lense-Thirring effect, Kerr metric [24]). Hence, only 
intrinsic spin can be connected with torsion.

The analogy with isospin suggests spin to be not simply a quantum number, but 
also the source of a gauge field. Like spin, isospin is described by 
the group $SU(2)$ [25]. 
When Heisenberg [25] introduced  
isospin, he supposed the (weak) nuclear force to be an exchange interaction, 
analogous to the spin exchange interaction with which he and Bethe were 
able to explain ferromagnetism and antiferromagnetism [26, 27]. 
Later, the Weinberg-Salam theory [10, 11] has shown isospin to be not simply 
a quantum number, but also the source of the 
weak nuclear interaction. 

The presented arguments suggest a gauge theory of gravity which requires 
a gauge boson of spin three that is associated with both torsion and 
intrinsic spin. 

Various gauge theories of gravity including either massless or massive 
torsion fields have been suggested (for a review and a detailed reference 
list see Ref. [18]). We will now argue for a non-zero rest mass of the 
tordion.

(i) According to gauge theories charge is conserved if and only if the 
rest mass of the associated gauge boson is exactly zero.
In contrast to total angular momentum, which is the sum of intrinsic 
spin and orbital angular momentum, intrinsic spin alone is not conserved. 
Hence, the tordion has to be massive.

(ii) Accelerated charges radiate. In rationalized units the spin $\hbar /2$ 
of an electron is greater than its electric charge $e$. If a tordion 
were massless, then the ``torsional'' part of the synchrotron radiation 
emitted by the electron would be stronger than its electromagnetic part. 
This would result in a significant difference between the theoretical 
(according to the standard model) and the actual energy of electrons 
after acceleration. Such a difference, were it real, is unlikely to 
have escaped discovery in particle accelerators.

(iii) According to Dirac [21], the electric-magnetic duality (i. e. the 
introduction of magnetic charges) yields quantized electric and magnetic 
charges. This result, however, is correct if and only if the 
electromagnetic field (i. e. both photon and magnetic photon) is massless. 
By contrast, the spin-mass duality introduced by Kibble [6] and Sciama 
[7, 8] does not yield quantized charges. Gravitational mass is not 
quantized. In the linearized approximation of general relativity a massive 
graviton would change deflection of light by the sun to $3/4$ its 
Einstein (and observed) value [28 - 30] (also Refs. [31, 32]). Hence, 
to agree spin-mass duality and massless graviton with non-quantized mass, 
we have to assume the tordion to be the massive gauge boson.

(iv) In rationalized units both Fermi's constant [33] of $V-A$ theory 
[34 - 36] and Newton's constant have dimension two. In Weinberg-Salam 
theory [10, 11], Fermi's constant turns out to be, up to a constant of 
order unity, the dimensionless coupling constant times the square of the 
inverse W-boson rest mass. By contrast, Newton's constant equals the 
square of the inverse Planck mass which, however, is not the rest mass 
of the (massless) graviton. A possibility is to interpret the Planck 
mass as the rest mass of the second gauge boson of gravity, the tordion. 

To conclude, the quantum field theory of gravity is presumably a gauge 
theory whose underlying group is the Poincar\'e group. This theory is 
supposed to include a massive torsion (and associated intrinsic spin) 
field which breaks the gauge invariance (spontaneously?). The Lagrangian 
is expected to have the dimension $-4$ and the coupling constant should 
be dimensionless. Finally, the classical, low energy limit has to regain  
general relativity.

\end{document}